\documentclass[iop,apjl]{emulateapj}
\usepackage{amssymb}		 
\usepackage{amsmath, amsthm, amssymb,amsfonts}
\usepackage[english]{babel}
\usepackage{epstopdf} 
\usepackage{color}
\usepackage{lipsum}
\usepackage{color, graphicx}

\newcommand\uv{{\bf u}}

\newcommand\ev{{\bf e}}

\newcommand\vv{{\bf v}}

\newcommand\xv{{\bf x}}

\newcommand\Bv{{\bf B}}

\newcommand\Ev{{\bf E}}

\newcommand\Jv{{\bf J}}

\newcommand\betai{\beta_{\rm i}}

\newcommand\bnabla{\boldsymbol{\nabla}}

\begin{document}

\title{Kinetic cascade in solar-wind turbulence: 3D3V hybrid-kinetic simulations with electron inertia}
\author{Silvio~Sergio~Cerri$^1$}
\author{Sergio~Servidio$^2$}
\author{Francesco~Califano$^1$}
\affiliation{$^1$Dipartimento di Fisica ``E. Fermi'', Universit\`a di Pisa, Largo B. Pontecorvo 3, 56127 Pisa, Italy}
\affiliation{$^2$Dipartimento di Fisica, Universit\`a della Calabria, 87036 Rende (CS), Italy}
\email{silvio.sergio@df.unipi.it}

\begin{abstract}

Understanding the nature of the turbulent fluctuations below the ion gyroradius 
in solar-wind turbulence is a great challenge. 
Recent studies have been mostly in favor of kinetic Alfv\'en wave (KAW) type of fluctuations,
but other kinds of fluctuations with characteristics typical of magnetosonic, whistler and 
ion Bernstein modes, could also play a role depending on the plasma parameters.
Here we investigate the properties of the sub-proton-scale cascade 
with high-resolution hybrid-kinetic simulations of freely-decaying turbulence
in 3D3V phase space, including electron inertia effects.
Two proton plasma beta are explored: the ``intermediate'' $\beta_p=1$ and ``low''
$\beta_p=0.2$ regimes, both typically observed in solar wind and corona.
The magnetic energy spectum exhibits $k_\perp^{-8/3}$ and $k_\|^{-7/2}$
power laws at $\beta_p=1$, while they are slightly steeper at $\beta_p=0.2$.
Nevertheless, both regimes develop a spectral anisotropy consistent with
$k_\|\sim k_\perp^{2/3}$ at $k_\perp\rho_p>1$, and pronounced small-scale intermittency.
In this context, we find that the kinetic-scale cascade is dominated by KAW-like 
fluctuations at $\beta_p=1$, whereas the low-$\beta$ case presents a more
complex scenario suggesting the simultaneous presence of different types of fluctuations.
In both regimes, however, a possible role of ion Bernstein type of fluctuations at the 
smallest scales cannot be excluded.

\end{abstract}


\maketitle

\section{Introduction}\label{sec:intro}

Nearly all astrophysical and space plasmas are in a turbulent state. 
In this context, the solar wind (SW) represents an ideal environment
for studying collisionless plasma turbulence from the magnetohydrodynamic (MHD) 
range down to kinetic scales~\citep{BrunoCarboneLRSP2013,ChenJPP2016}.
Increasingly accurate in-situ measurements of SW turbulence 
down to electron scales have been available over the past years~\citep{BalePRL2005,%
AlexandrovaAPJ2008,SahraouiPRL2010,HeAPJL2012,%
RobertsAPJ2013,ChenPRL2013},
showing the presence of breaks in the electromagnetic fluctuations at kinetic scales.
In the proton kinetic range, for instance, the typical slope for the magnetic energy spectrum
is found to be between $-2.5$ and $-3$, i.e., steeper than the correspondent spectrum at MHD scales,
while the electric spectrum becomes simultaneously shallower below the proton gyroradius scale. 
A wide number of theoretical models~\citep{VainshteinJETP1973,GaltierPOP2003,%
ChoLazarianAPJL2004,HowesJGR2008,SchekochihinAPJS2009,%
BoldyrevAPJL2012,BoldyrevAPJ2013,PassotSulemAPJL2015}
and numerical investigations~\citep{ShaikhMNRAS2009b,%
ValentiniPRL2010,HowesPRL2011,ServidioPRL2012,ServidioAPJL2014,%
ToldPRL2015,SulemAPJ2016,FranciAPJ2015,FranciAPJ2016,CerriAPJL2016,Groselj2017} 
have been exploited in order to explain the observed behavior of SW turbulent spectra, 
mostly in terms of the properties of fluctuations derived from wave physics.
In this context, the observed spectra at kinetic scales are usually interpreted as 
a cascade of kinetic Alfv\'en waves (KAWs) and/or of higher frequency waves, 
such as magnetosonic (MS), whistler waves (WWs) and/or ion Bernstein (IB) modes. 
Most of the SW observations points towards a cascade of KAW-like fluctuations 
at $\beta\sim1$~\citep{SahraouiPRL2010,HeAPJL2012,RobertsAPJ2013,ChenPRL2013},
where $\beta$ is the ratio between thermal and magnetic pressures, 
although also whistler-like turbulence have been observed~\citep{NaritaGRL2011,NaritaAPJL2016}. 
In fact, theoretical arguments suggest that different kinds of fluctuations could coexist and interact,
depending on the plasma parameters~\citep{StawickiJGR2001,GaryJGR2009,MithaiwalaPOP2012,PodestaJGR2012}. 
This idea has been recently explored via 2D numerical simulations that suggested
an increasingly KAW-like turbulence as $\beta$ increases, 
whereas a more complex scenario - i.e., a mixture of different kind of fluctuations, including KAW-like ones - 
seems to emerge in the low-$\beta$ regimes~\citep{CerriAPJL2016,CerriJPP2017,Groselj2017}.
However, interpreting the turbulent cascade only in terms of wave physics is perhaps limiting
and unsatisfactory~\citep{MatthaeusAPJ2014}.
Recently, the idea that magnetic reconnection can play a fundamental role 
in the formation of the small-scale spectrum has emerged~\citep{CerriCalifanoNJP2017,%
Mallet2017b,LoureiroBoldyrev2017b,Franci2017}.
These interpretations are somewhat at odds with the picture of turbulence 
made solely by a cascade of waves, as pointed out also by the intermittent behavior 
of SW turbulence~\citep{SorrisoValvoGRL1999,PerriPRL2012,KiyaniAPJ2013,OsmanPRL2014}. 

In this Letter, we present high-resolution 3D3V simulations
of the turbulent cascade below the proton gyroradius within
a hybrid Vlasov-Maxwell (HVM) model of plasma 
including finite electron inertia ($m_p/m_e=100$).
Here, we focus on the spectral and intermittent properties of kinetic-scale turbulence 
in order to address the question of a possible dependence 
of the physics of such cascade on the plasma beta parameter. 
We remind that our hybrid approach, although not retaining all the electron kinetic effects, 
fully captures the ion kinetic physics and allows for both KAWs, 
magnetosonic, whistlers and ion Bernstein fluctuations to be present.
We want to stress that here we analyze the properties of the turbulent fluctuations 
and we relate them to the characteristic features of the corresponding linear modes, 
but in doing this we are not assuming that turbulence is made by a sea of linear waves:
the aim of the analysis is to understand and classify the characteristics of turbulent fluctuations
in analogy with those derived via linear theory.

\section{The HVM model and simulations setup}

The HVM model couples fully-kinetic protons to fluid electrons 
through a generalized Ohm's law~\citep{MangeneyJCP2002,ValentiniJCP2007}. 
The model equations, normalized with respect to the proton characteristic quantities 
(mass $m_p$, gyrofrequency $\Omega_p$ and inertial length, $d_p$) and to the Alfv\'en speed $v_A$, read
\begin{equation}\label{eq:HVM_Vlasov}
\frac{\partial\,f}{\partial t}\, +\, \vv\cdot\frac{\partial\,f}{\partial\xv}\, +\,
 (\Ev+\vv\times\Bv)\cdot\frac{\partial\,f}{\partial\vv}\, =\, 0\,,
\end{equation}
\begin{equation}\label{eq:HVM_Ohm}
(1-d_e^2\nabla_\perp^2)\Ev \,=\, -\,\uv\times\Bv \,+\,
 \frac{\Jv\times\Bv}{n} \,-\,\frac{\bnabla p_e}{n}\,,
\end{equation}
\begin{equation}\label{eq:HVM_Maxwell}
\frac{\partial\,\Bv}{\partial t}\,=\, -\,\bnabla\times\Ev\,,\quad
  \bnabla\times\Bv\, =\, \Jv\,
\end{equation}
where $f(\xv,\vv,t)$ is the proton distribution function, $d_e^2 = m_e$ is the electron skin depth,
quasi-neutrality $n_p\simeq n_e\equiv n$ is assumed, 
and the displacement current is neglected in the Amp\'ere's law. 
In the generalized Ohm's law, the leading electron inertia term 
$d_e^2\nabla^2 \simeq d_e^2\nabla_\perp^2$ has been included
(assuming $k_\|^2 \ll k_\perp^2$ and a naturally anisotropic cascade).
An isothermal closure for the electron pressure,
$p_e=nT_{0,e}$, is adopted, and number density, $n$, 
and proton mean velocity, $\uv$, are computed as $v$-space moments
of $f$. 

We initialize the simulations with a Maxwellian 
proton distribution function with isotropic temperature $T_{0,p}$
and an electron fluid with $T_{0,e}=T_{0,p}$, 
embedded in a uniform background magnetic field
$\Bv_0=B_0\ev_z$ with $B_0=1$. 
We further impose initial random large-scale 3D 
isotropic magnetic perturbations, $\Bv=\Bv_0+\delta\Bv$, with wave numbers
$0.1\leq kd_i\leq0.5$ and $\delta B^{\rm rms}\simeq0.23$.
We use $384^2$ grid points in the perpendicular $xy$-plane
and $64$ points in the parallel $z$ direction, uniformly distributed
to discretize a periodic simulation box with $L_\perp = 10\,\pi\,d_p$ and
$L_\|=2L_\perp=20\,\pi\,d_p$, corresponding to a perpendicular
resolution $\Delta x=\Delta y \simeq 0.08\,d_p = 0.8\,d_e$ 
and $\Delta z\simeq d_p$. 
This corresponds to a spectral domain that spans more than two decades 
in perpendicular wave numbers, $0.2\leq k_\perp d_p\leq38.4$, 
and more that one decade in its parallel counterpart, $0.1\leq k_\| d_p\leq3.2$. 
We apply (weak) spectral filters during the simulation
in order to prevent spurious numerical effects at the smallest scales~\citep{LeleJCP1992},
thus determining a cut-off in the turbulent energy spectra 
for $k_\perp d_p\gtrsim20$ and for $k_\|d_p\gtrsim2$. 
The velocity domain is limited in each direction 
by $v_{\rm max} = \pm\,5\,v_{{\rm th,p}}$  for the $\beta_p=1$ case
and by $v_{\rm max} = \pm\,8\,v_{{\rm th,p}}$ for $\beta_p=0.2$, 
with $51^3$ and $61^3$ uniformly distributed velocity grid points, respectively. 

\section{Anisotropy and intermittency of kinetic turbulence}

\begin{figure}[!t]
 \flushleft
 \includegraphics[width=0.5\textwidth]{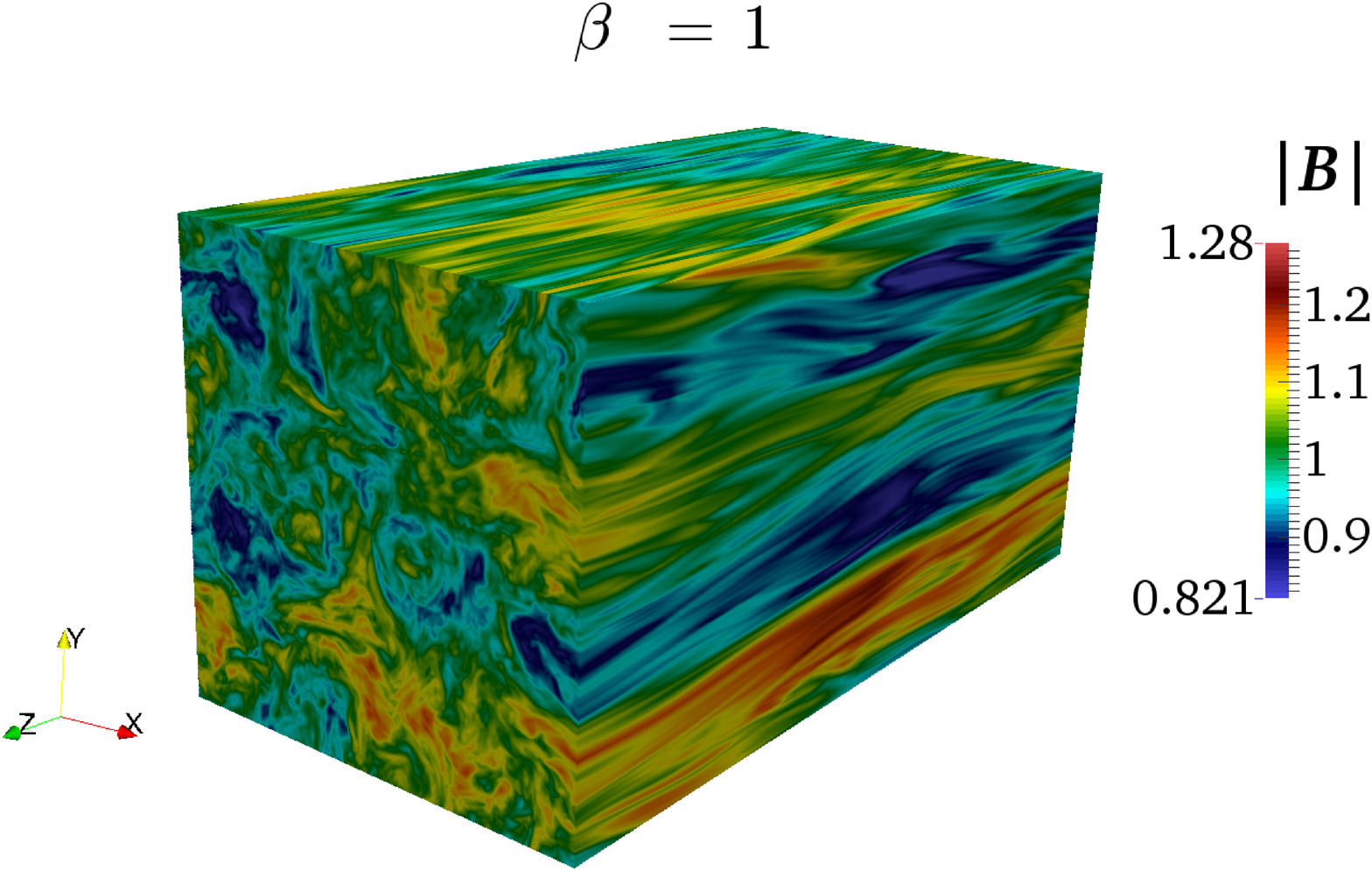}
 \includegraphics[width=0.5\textwidth]{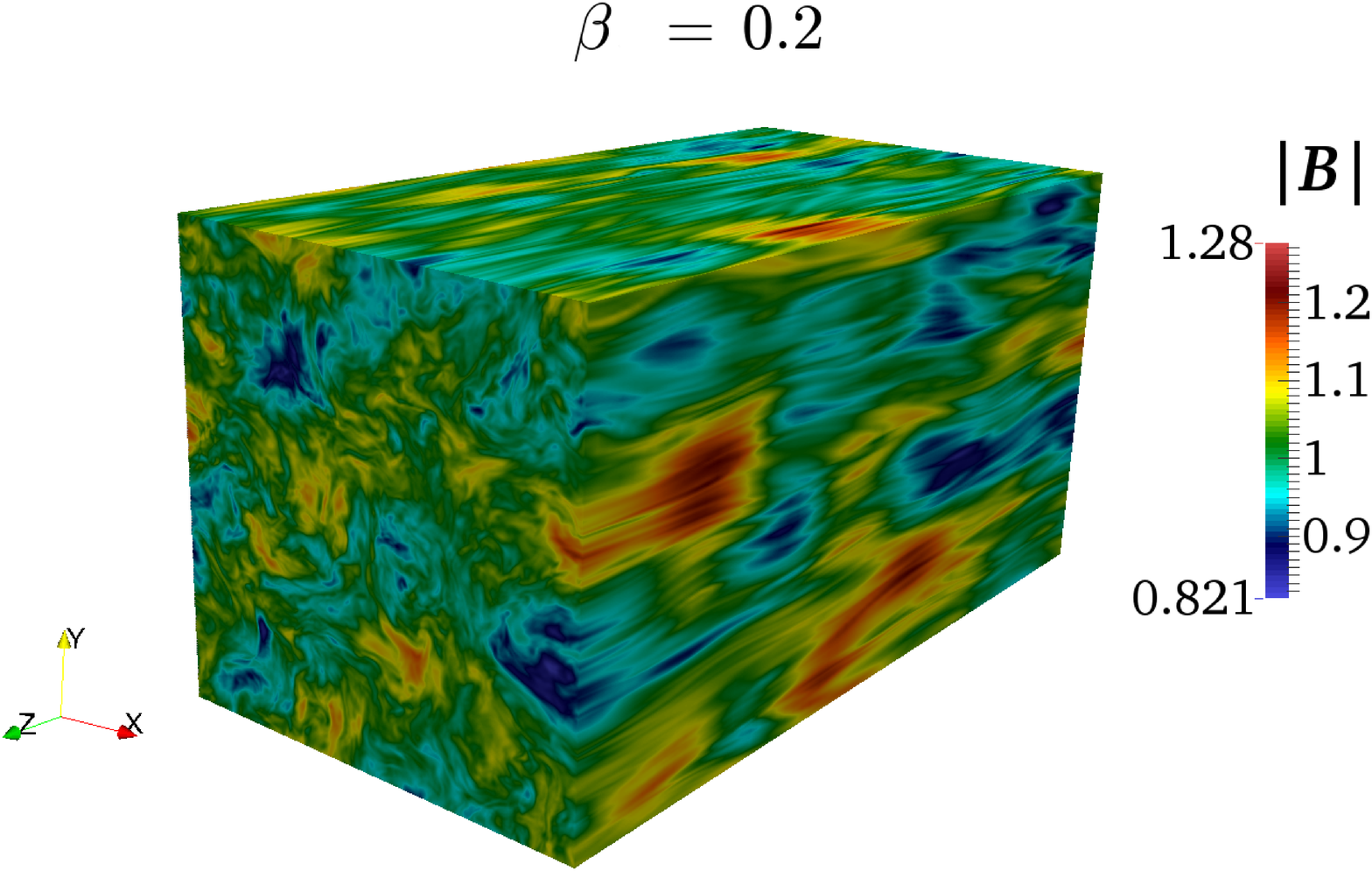}
 \caption{Three-dimensional representation of the magnetic field magnitude, $|\Bv|$, in the fully-developed turbulent state for $\beta_p=1$ and $\beta_p=0.2$ regimes (top and bottom panel, respectively).}
 \label{fig:B3D}
\end{figure} 
Within a few outer-scale nonlinear times the initial condition freely-decays 
into a fully-developed turbulent state at $t= t_*$.
Such time is identified by a peak in the 
root-mean-square current density, $J^{\rm rms}$.
In order to increase the statistics, the spectral analysis 
of turbulent fluctuations presented here includes 
a short time average over $\Delta t=10\,\Omega_p^{-1}\ll t_*$, starting from $t_*$. 

Before discussing the spectral properties,
a difference between the $\beta_p=1$ and $\beta_p=0.2$ regimes
is first pointed out at the level of the spatial structures emerging 
in the fully-developed turbulent state.
This is shown in Fig.~\ref{fig:B3D} where we draw the 
three-dimensional contours of the magnetic field magnitude 
at $t=t_*$ in the two distinct regimes (top and bottom panel 
for $\beta_p=1$ and $\beta_p=0.2$, respectively).
As expected, starting with the same initially isotropic condition, 
in both cases the fluctuations gradually cascades into strongly anisotropic turbulence.
However, while the $\beta_p=1$ regime exhibits perpendicular small-scale structures
and very elongated fluctuations along $\Bv_0$ that are typical of Alfv\'enic turbulence, 
the $\beta_p=0.2$ case presents shorter parallel structures 
that are instead reminiscent of magnetosonic fluctuations.

\begin{figure}[!t]
  \flushleft\includegraphics[width=0.5\textwidth]{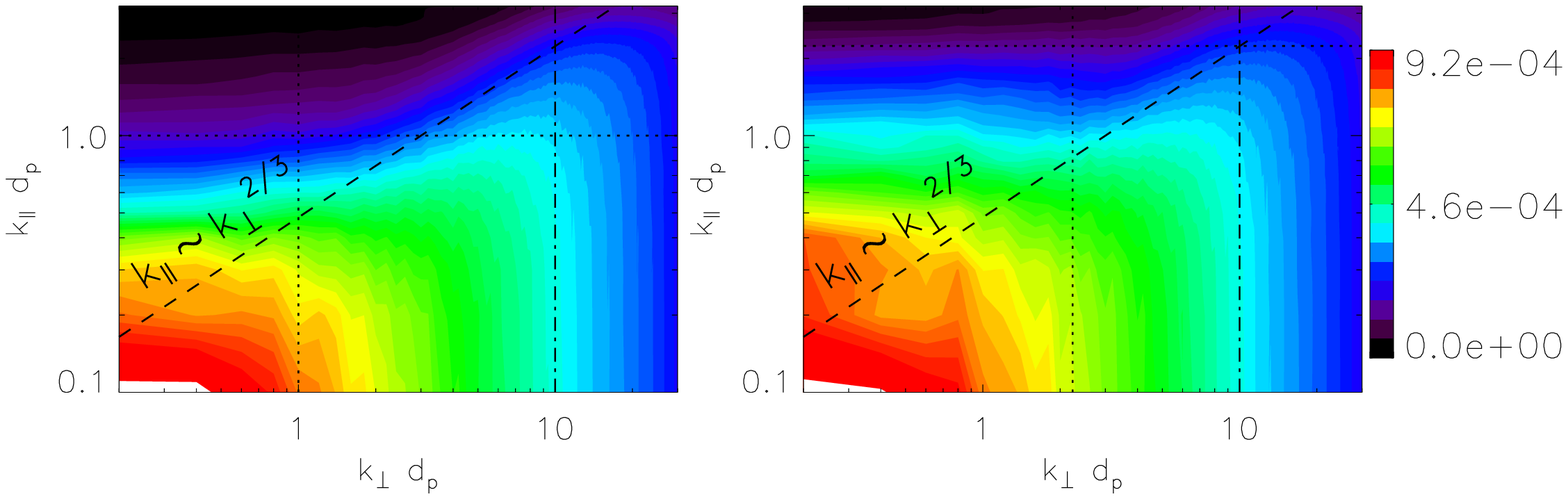}
  \vspace{-0.75cm}
  \flushleft\includegraphics[width=0.5\textwidth]{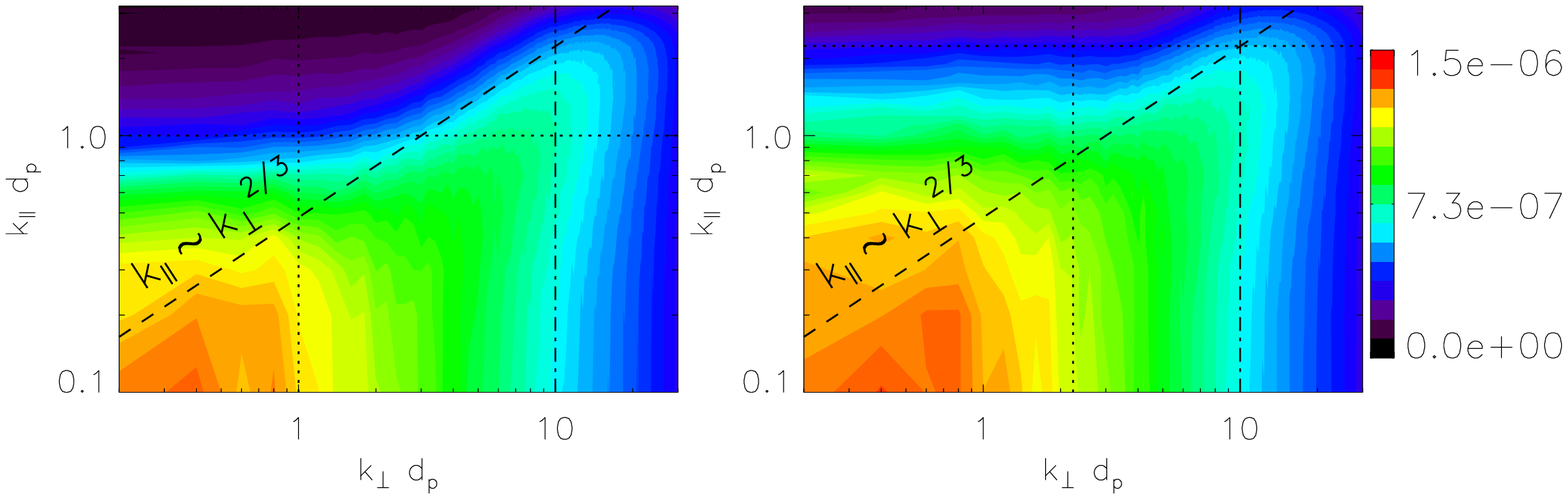}
 \caption{Two-dimensional energy spectrum in the ($k_\perp$, $k_\|$) plane of the total magnetic fluctuations and of the parallel electric fluctuations, ${\cal E}_{B}(k_\perp,k_\|)$ and ${\cal E}_{E\|}(k_\perp,k_\|)$ (top and bottom row, respectively) for $\betai=1$ (left column) and $\beta_i=0.2$ (right column).}
 \label{fig:Spectra2D}
\end{figure} 

The spectral anisotropy of the turbulent fluctuations is shown in Fig.~\ref{fig:Spectra2D}, 
where we draw the two-dimensional energy spectrum of the total magnetic fluctuations, $\delta B$ (top panels), 
and of the parallel electric fluctuations, $\delta E_\|$ (bottom panels), for both regimes 
(left and right column for $\beta_p=1$ and $\beta_p=0.2$, respectively).
Anisotropy is observed also at $k\rho_p<1$,
although this region contains few $k$ points and is thus less relevant.
At smaller scales, $k_\perp\rho_p>1$, the anisotropy is instead evident: 
the turbulent cascade is mainly perpendicular to $\Bv_0$ 
and the fluctuations seem to follow a $k_\|\sim k_\perp^{2/3}$ pattern.
This is more pronounced in the $\beta_p=1$ case, where the available sub-proton-scale 
range is larger than in the low-$\beta_p$ counterpart. 
Such pattern reveals a weaker anisotropy than the $k_\|\sim k_\perp^{1/3}$
scaling phenomenologically expected for both KAW and whistler 
turbulence~\citep{ChoLazarianAPJL2004,SchekochihinAPJS2009}, 
and it is rather in agreement with the one predicted for turbulence
mainly concentrated within 2D sheet-like structures~\citep{BoldyrevAPJL2012}.
The spectra of fluctuations in the other quantities show the same behavior (not shown here).

A classical intermittency analysis has been performed on both simulations 
at about the peak of the nonlinear activity. 
In order to define the large scale limit of the inertial range, we evaluated the 
perpendicular and parallel auto-correlation functions, respectively defined as 
$C(r_\perp) = \langle \delta\Bv(\xv+{\bf r}_\perp)\cdot\delta\Bv(\xv) \rangle$ and 
$C(r_\|) = \langle \delta\Bv(\xv+{\bf r}_\|)\cdot\delta\Bv(\xv) \rangle$~\citep{Frisch95}. 
We assumed isotropy in the perpendicular $xy$-plane, with the parallel direction along $\Bv_0$, i.e. along~$z$. 
The $e$-folding length gives approximately the integral scale which is about $\lambda_\perp\sim 3 d_p$ 
in the perpendicular direction (corresponding to $k_\perp d_p\sim2$), for both regimes. 
The situation is different in the parallel direction, where the parallel correlation length is $\lambda_\|\sim8 d_p$ for $\beta_p=0.2$, 
while is $\lambda_\|\sim12 d_p$ for $\beta_p=1$ (corresponding to $k_\|d_p\sim0.8$ and $\sim0.5$, respectively). 
This is in qualitative agreement with the features spotted in Fig.~\ref{fig:B3D}, 
and, quantitatively, with the corresponding spectra (see Fig.~\ref{fig:Spectra1D}), 
indicating differences already in the large-scale properties of the fluctuations 
possibly due to a different decorrelation mechanism along the mean field.

\begin{figure}[!t]
\flushright\includegraphics[width=0.48\textwidth]{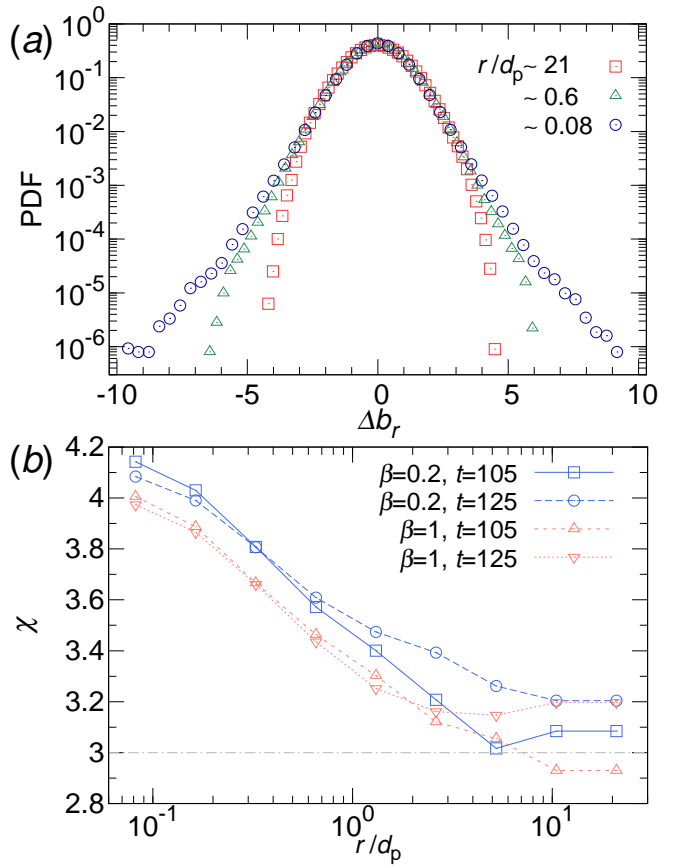}
\caption{Top: PDFs of magnetic increments for $\beta_p=0.2$, at different perpendicular lags. Bottom: Scale dependent 
kurtosis $\chi$. Comparison of the scale-dependent Kurtosis, at different times, for the two betas.}
\label{fig:pdfs}
\end{figure}
The level of intermittency can be better quantified by the PDFs 
of the magnetic field increments at a given scale $r$, defined as 
\begin{equation}
\Delta b_r \equiv  [\delta\Bv(\xv+{\bf r})-\delta\Bv(\xv)]\cdot \hat{{\bm r}}. 
\label{eq:dpr}
\end{equation}
We show here the statistics of the perpendicular increments, namely $r\equiv r_\perp$, 
spanning this increment from lengths larger than the correlation scale $\lambda_\perp$,
down to the smallest available scale ($\Delta x\sim0.08\,d_p$). 
These distributions are reported in Fig.~\ref{fig:pdfs}-(a) for the $\beta_p=0.2$ regime 
for three cases, namely $r/d_p=21, 0.6$ and $0.08$. 
These PDFs, as expected, become increasingly intermittent going towards smaller scales. 
In order to compare among cases, and among different times, we measured the scale-depended kurtosis $\chi$ 
-- the fourth-order moment of the increments in Eq. (\ref{eq:dpr}) -- that can be measured as
\begin{equation}
\chi = \frac{\langle \Delta b_r^4 \rangle}{\langle \Delta b_r^2\rangle^2}.
\label{eq:chi}
\end{equation}
This quantity is reported in Fig.~\ref{fig:pdfs}-(b), as a function of the perpendicular scale $r$, 
for the two values of $\beta$, at two distinct times.
At large scale, for $r>5 d_p$, the distribution becomes Gaussian, where $\chi\sim3$, 
in agreement with the computation of the correlation lengths. 
At small scales, in the inertial range of turbulence, there is an enhancement due to the 
intermittent nature of the cascade, due to the presence of coherent structures and non-linear waves. 
At the smallest scales, a saturation of the multifractality is observed, in agreement with observations in the solar wind. 
In fact the study of high-order structure functions up to the 6$^{th}$ moment and of their exponents, 
shows deviation from monofractality (not shown here).
Here this process of saturation might also be slightly affected by the presence of artificial dissipation. 
It is important to notice, that at scales in the inertial-dispersive range, 
the case with $\beta_p=0.2$ is more intermittent than the $\beta_p=1$ regime, 
indicating a higher degree of coherency in the small-scale fluctuations.

\section{Spectral features of kinetic-scale fluctuations}

\begin{figure}[!t]
  \vspace{-.2cm}
  \flushleft\hspace{-0.8cm}\includegraphics[width=0.52\textwidth]{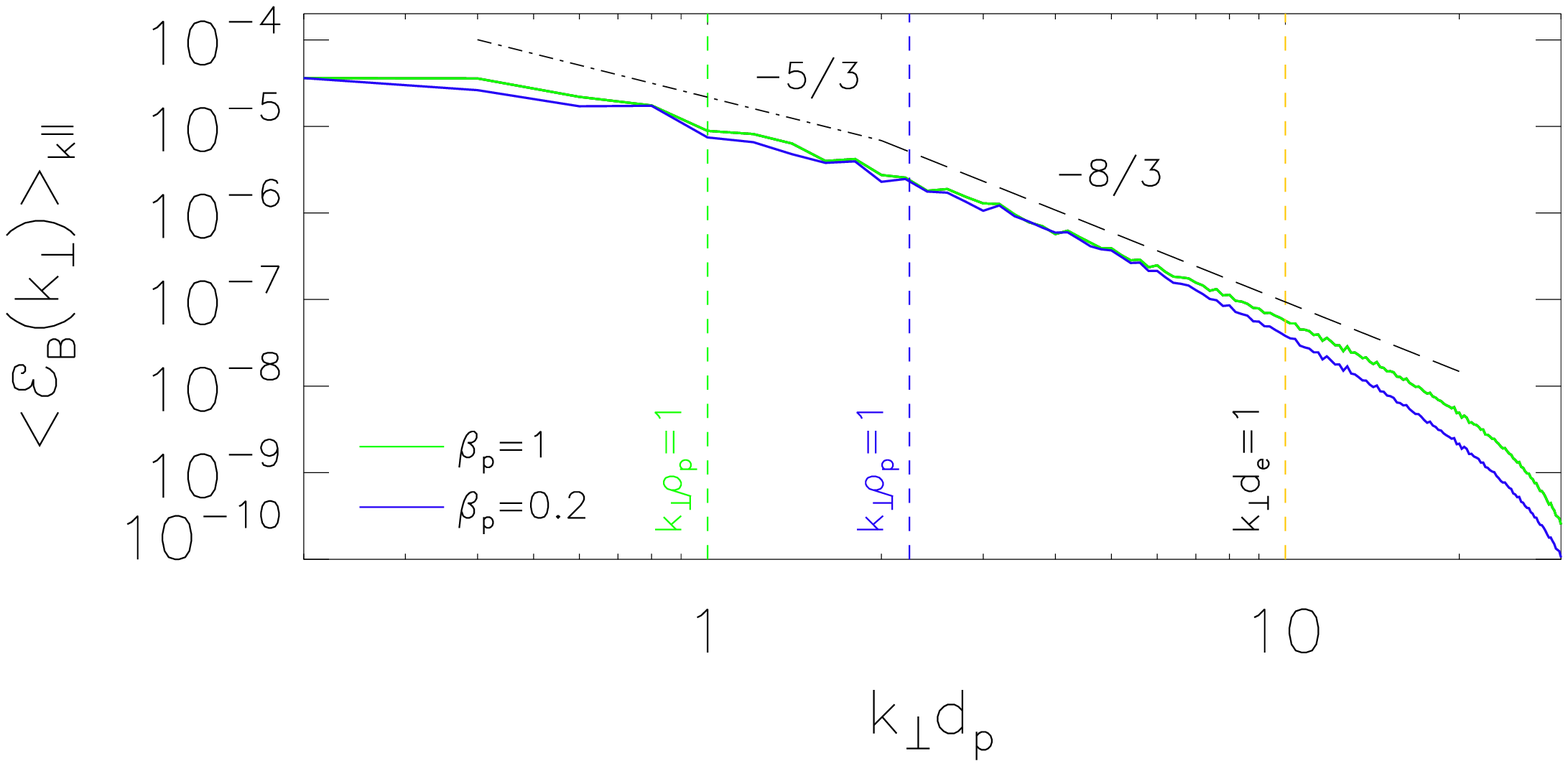}
  \vspace{-1.2cm}
  \flushleft\hspace{-0.8cm}\includegraphics[width=0.52\textwidth]{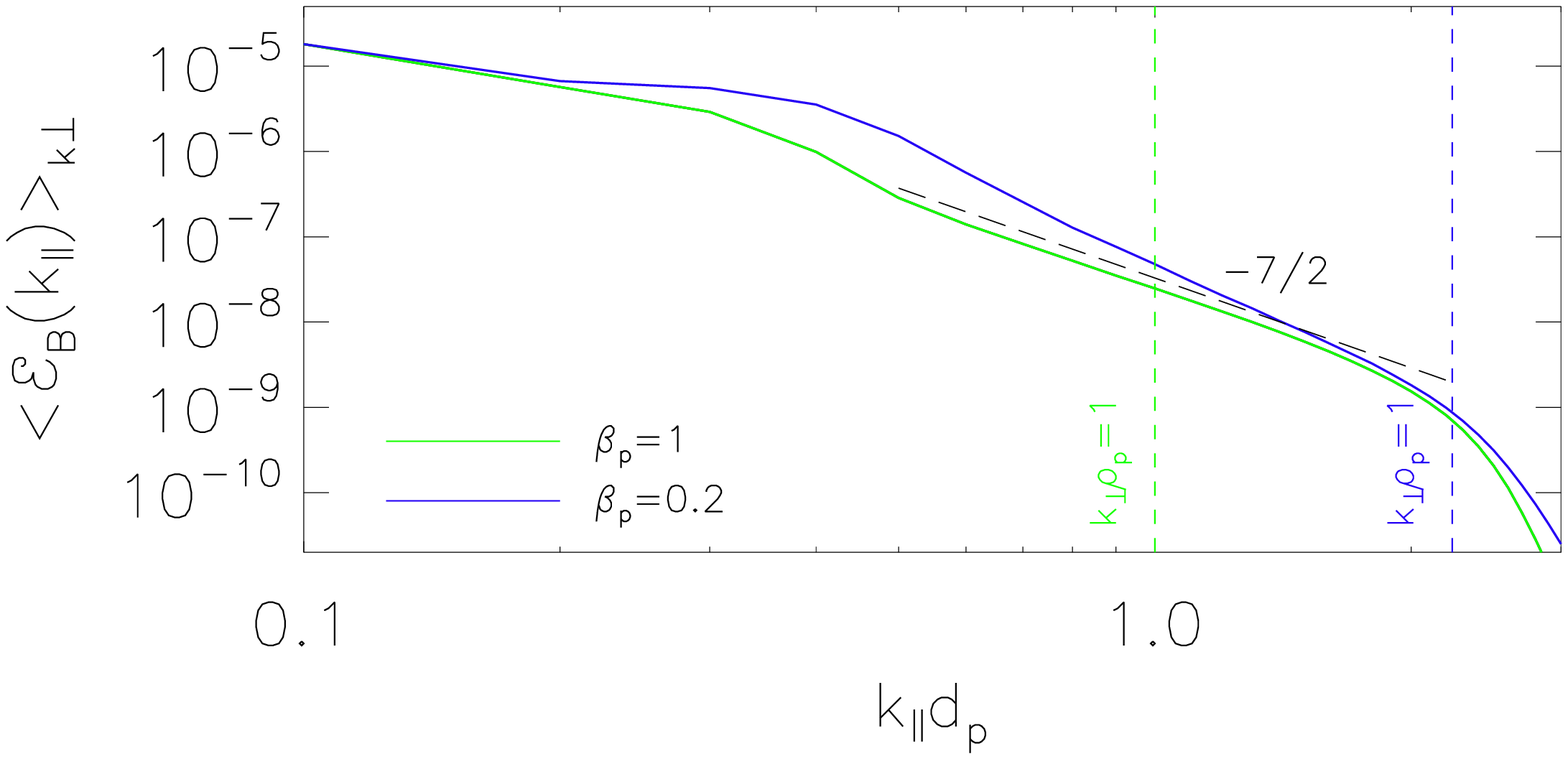}
  \vspace{-0.55cm}
 \caption{Total magnetic energy spectrum versus $k_\perp$ (top panel) and versus $k_\|$ (bottom panel), for $\betai=0.2$ and $\beta_i=1$ (blue and green line, respectively).}
 \label{fig:Spectra1D}
\end{figure} 

In Fig.~\ref{fig:Spectra1D}, we show the one-dimensional magnetic energy spectrum
for both regimes (green and blue line for $\beta_p=1$ and $\beta_p=0.2$, respectively):
the $k_\|$-averaged spectrum versus $k_\perp$, $\langle{\cal E}_B(k_\perp)\rangle_{k\|}$ (top frame), 
and the $k_\perp$-averaged counterpart versus $k_\|$, $\langle{\cal E}_B(k_\|)\rangle_{k\perp}$ (bottom frame). 
The average procedure, e.g. $\langle{\cal E}(k_\perp)\rangle_{k\|}$, here is defined as the summation 
of ${\cal E}(k_{\|,i},k_{\perp,j})$ over the points of the $\{k_{\|,i}\}_{i=1,\dots,N_\|}$ grid, divided by those number of points, $N_\|$. 
Such procedure, when specified, can be restricted to a $k_\perp$-dependent sub-set of points, $n_\|(k_\perp)$, 
of the entire $k_\|$ grid (see later in this Section).
At large perpendicular scales, $0.4\lesssim k_\perp \rho_p\lesssim2$, a nearly $-5/3$ power law is visible
in both cases, although the MHD range is too limited to draw conclusions.
At small perpendicular scales, $k_\perp\rho_p\gtrsim2$, the $\beta_p=1$ regime exhibits 
a power law very consistent with a $-8/3$ slope 
(this has been verified through compensated spectra),
while the $\beta_p=0.2$ case shows a steeper spectrum, close to $k_\perp^{-3}$.
For small parallel wave numbers, roughly $k_\|\rho_p\lesssim0.5$, 
an excess of magnetic energy is present for $\beta_p=0.2$
and no clear power laws can be drawn for both regimes.
For $k_\|\rho_p\gtrsim0.5$, instead, a $-7/2$ slope is observed at $\beta_p=1$,
whereas at lower $\beta$ it is again steeper (roughly between $k_\perp^{-9/2}$ and $k_\perp^{-5}$).
Note that the kinetic-range cascade, expected to take place at $k\rho_p>1$, 
in the parallel wave numbers already starts at $k_\|\rho_p\sim0.5$ due to the 
anisotropic nature of the turbulent cascade itself (cf. Fig.~\ref{fig:Spectra2D}).
In particular, consistently with the spectral anisotropy and the intermittency analysis, 
the observed power laws for the magnetic spectrum at $\beta_p=1$, 
i.e., $\propto k_\perp^{-8/3}$ and $\propto k_\|^{-7/2}$, 
are in agreement with those predicted in \citet{BoldyrevAPJL2012}.
\begin{figure}[!t]
  \vspace{-.2cm}
  \flushleft\hspace{-0.8cm}\includegraphics[width=0.52\textwidth]{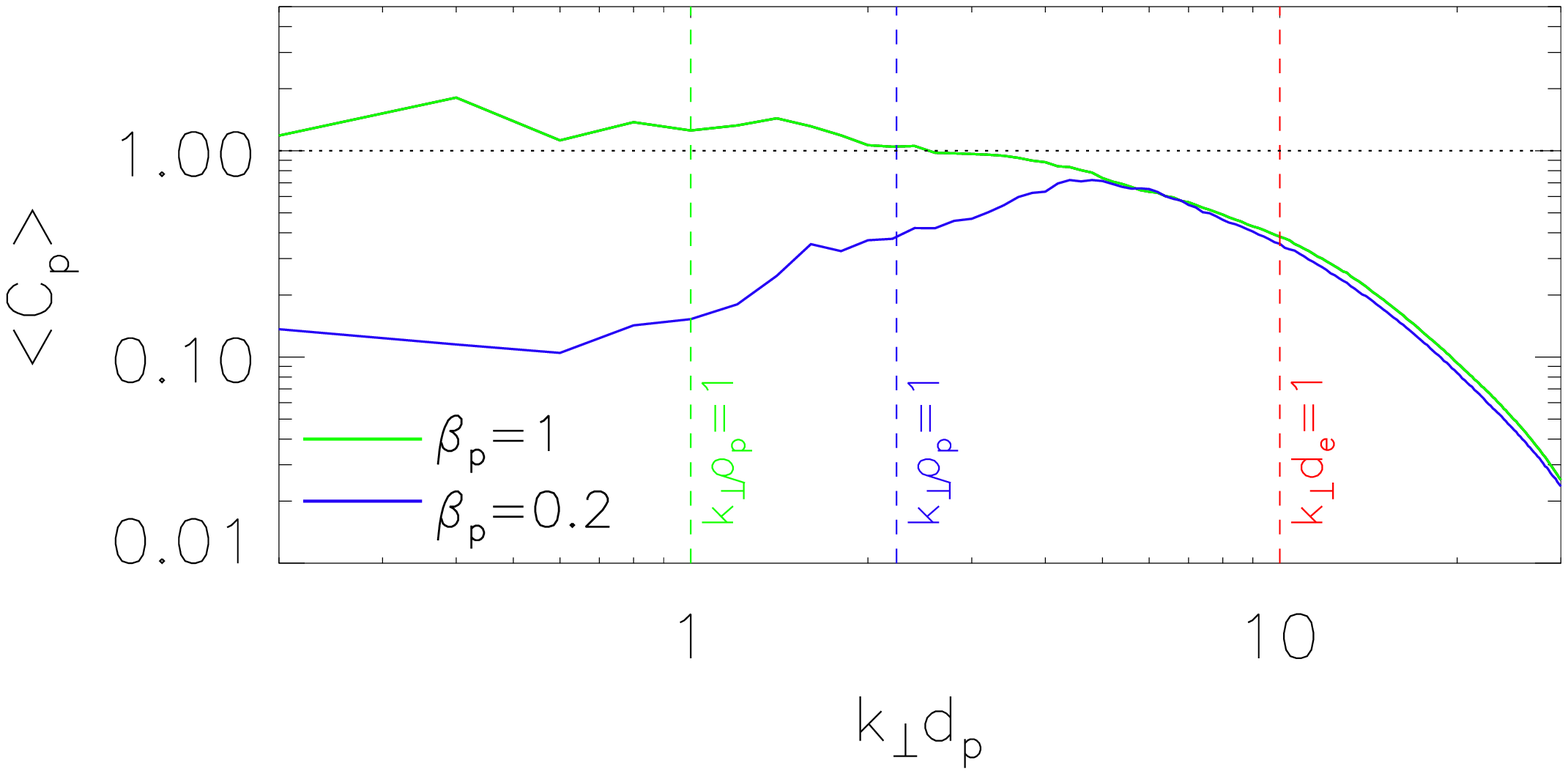}
  \vspace{-1.2cm}
  \flushleft\hspace{-0.8cm}\includegraphics[width=0.52\textwidth]{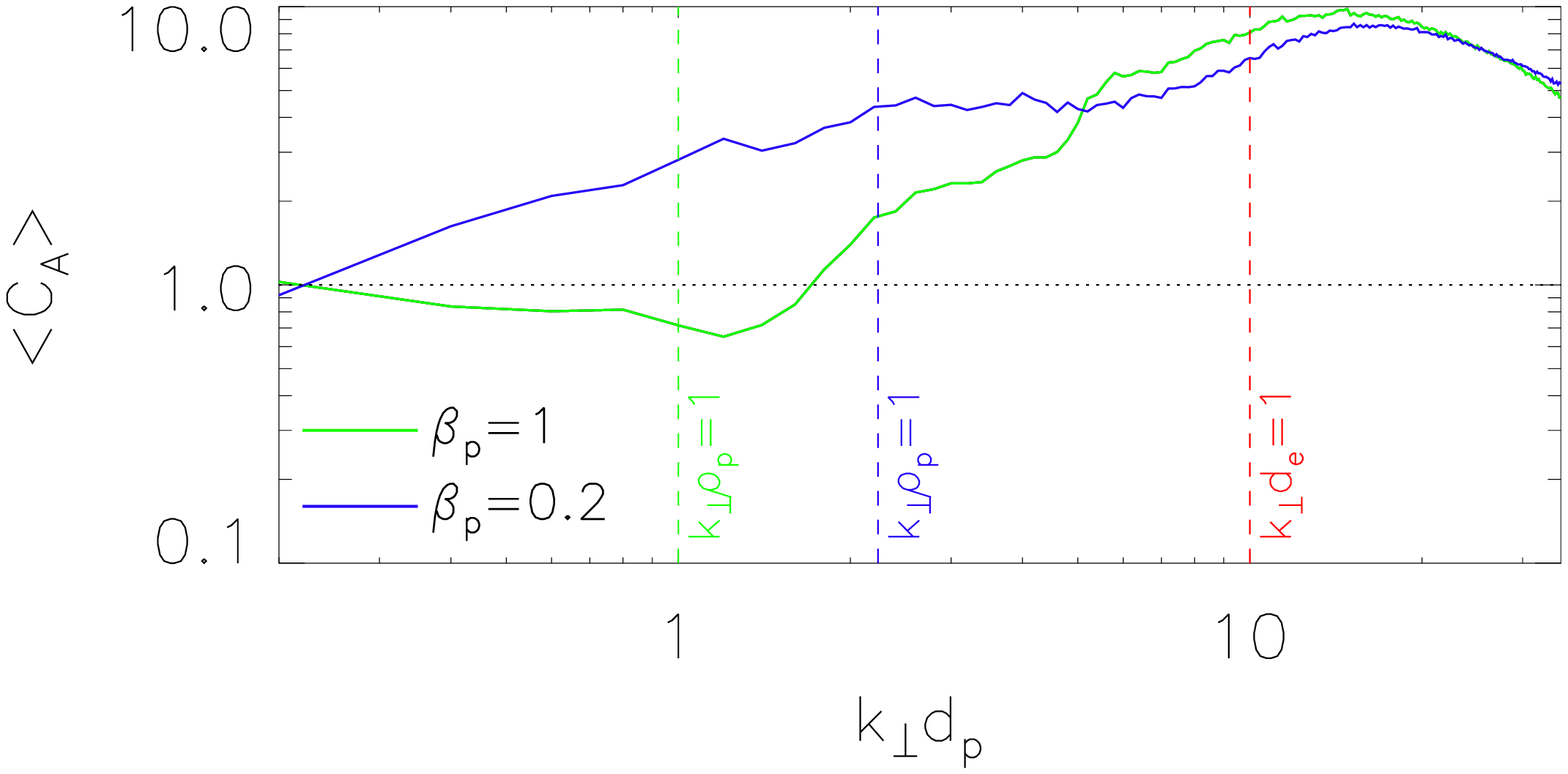}
  \vspace{-1.2cm}
  \flushleft\hspace{-0.8cm}\includegraphics[width=0.52\textwidth]{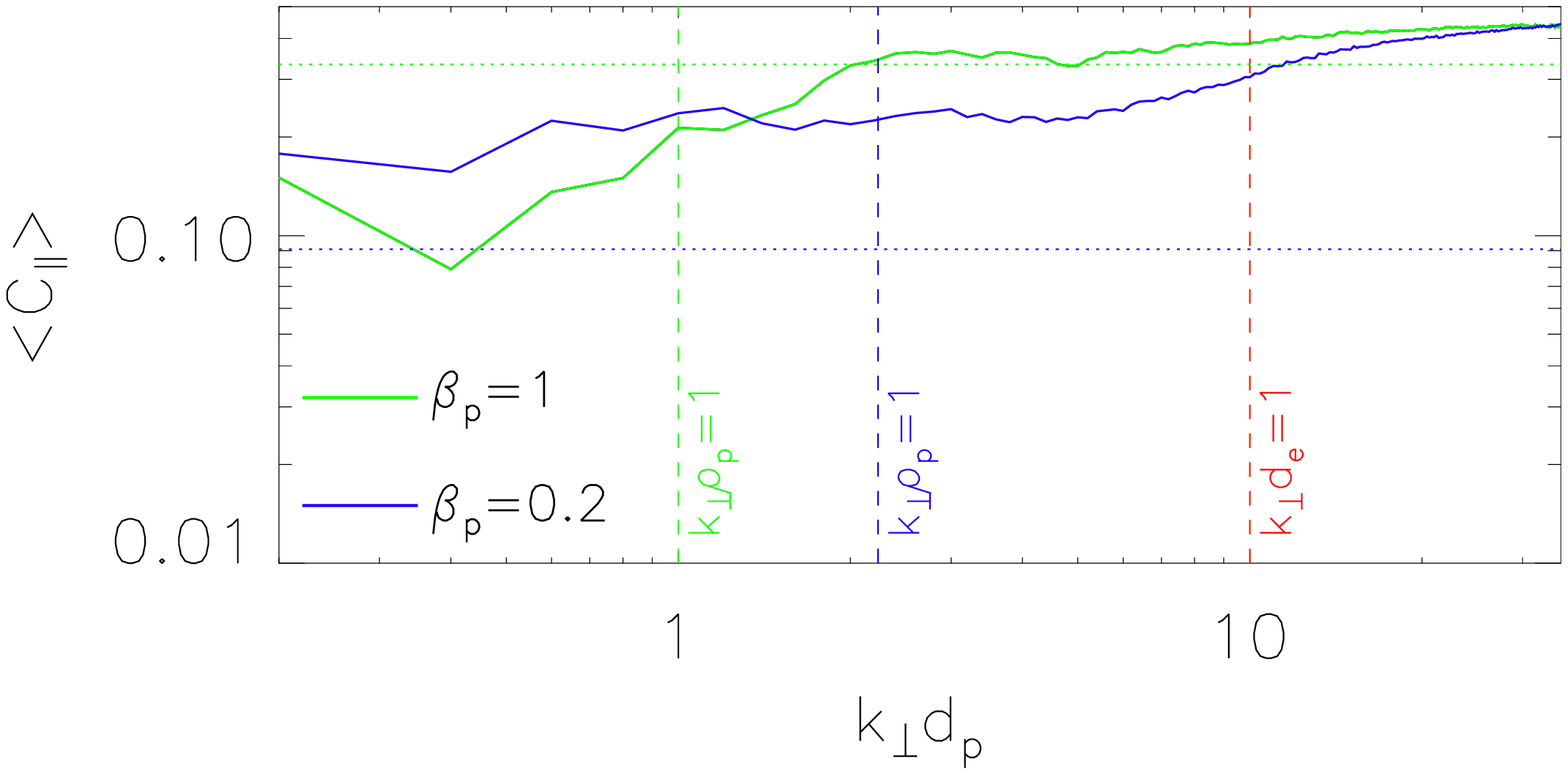}
  \vspace{-0.55cm}
 \caption{Averaged spectral ratios in Eq.~(\ref{eq:ratios}) versus $k_\perp$ for $\betai=0.2$ and $\beta_i=1$ (blue and green line, respectively). The average has been taken over those $k_\|$ such that $k_\|\leq k_\perp^{2/3}\rho_p^{-1/3}$ (cf. Fig.~\ref{fig:Spectra2D}).}
 \label{fig:SpectraRatio}
\end{figure} 

A useful tool for the investigation of turbulent fluctuations properties
are the spectral ratios of different quantities~\citep{ChenPRL2013,%
CerriAPJL2016,CerriJPP2017,ChenBoldyrev2017,HuangAPJL2017,Groselj2017}.
Here, in order to highlight the characteristic 
behavior of small-scale fluctuations in the two different regimes,
we consider the following quantities:
\begin{equation}\label{eq:ratios}
C_p\,\equiv\,\beta_p^2\,\frac{\delta n^2}{\delta B_\|^2}\,,\quad
C_A\,\equiv\,\frac{\delta E_\perp^2}{\delta B_\perp^2}\,,\quad
C_\|\,\equiv\,\frac{\delta B_\|^2}{\delta B^2}\,,
\end{equation}
where $\tau\equiv T_{0,e}/T_{0,i}=1$ has been already assumed in normalizing $C_p$.
Let us relate them to the characteristic signatures that the 
main oblique modes would leave on the above ratios~\citep{SchekochihinAPJS2009,BoldyrevAPJ2013}.
Since we are interested in the oblique fluctuations and given the
anisotropic behavior of the turbulent energy cascade shown in Fig.~\ref{fig:Spectra2D}, 
the ratios defined above will be averaged over parallel wave numbers 
such that $k_\|\leq k_\perp^{2/3}\rho_p^{-1/3}$.
The resulting ratios are thus function of $k_\perp$ only, 
highlighting the properties of the main turbulent fluctuations
and their connection with previous 2D numerical 
studies~\citep{CerriAPJL2016,CerriJPP2017,Groselj2017}.

We first consider $C_p$ (Fig.~\ref{fig:SpectraRatio}, top panel): 
the normalized ratio between density and parallel magnetic fluctuations 
is expected to be unity, $C_p\approx1$, for low-frequency Alfv\'enic/KAW fluctuations, 
whereas higher frequency modes such as MS, WWs and IB should leave this ratio much smaller, namely $C_p\ll1$.
For $\beta_p=1$, the $C_p$ ratio is about unity in nearly all the $k_\perp$ range,
which is a signature of turbulence dominated by low-frequency Alfv\'enic/KAW-like fluctuations.
In the $\beta_p=0.2$ case, instead, we obtain $C_p\ll1$ at large scales, $k_\perp\rho_p<1$, 
and it then increases for $k_\perp\rho_p>1$, reaching values similar to those observed at $\beta_p=1$.
In both regimes the behavior of $C_p$ at the smallest scales, $k_\perp d_p\gg1$, 
is most likely due to a combined effect of $k_\perp d_e$ terms~\citep{ChenBoldyrev2017} 
and by the enhanced coupling of the MS, WWs and KAWs with the ion Bernstein branches
in the oblique electromagnetic case~\citep{PodestaJGR2012}.

Second, we consider $C_A$ (Fig.~\ref{fig:SpectraRatio}, middle panel): 
at $k_\perp\rho_p<1$, this ratio is expected to be unity for Alfv\'enic fluctuations, $C_A\approx1$,
and to increase as $C_A~\simeq~\frac{1}{2}~\frac{\beta_p}{1+\beta_p}~(k_\perp\rho_i)^2$
for $k_\perp\rho_p>1$, i.e., in the KAW regime this ratio strongly depends on $\beta_p$. 
In the WWs regime, instead, this ratio does not depend on the beta and also increases
as $k_\perp^2$: $C_A\simeq2(k_\perp\rho_i)^2$. 
Qualitatively, the relation $C_A^{WW}\gtrsim  C_A^{KAW(\beta=1)}\gtrsim C_A^{KAW(\beta=0.2)}$ holds.
From Fig.~\ref{fig:SpectraRatio} (middle panel), the behavior of $C_A$ at $\beta_p=1$
is again consistent with predominantly Alfv\'enic/KAW-like fluctuations, whereas at $\beta_p=0.2$
the large-scale behavior is consistent with MS/WW-like fluctuations.
Nevertheless, due to the fact the the above qualitative relation $C_A^{(\beta=1)}\gtrsim C_A^{(\beta=0.2)}$ 
is recovered at high $k_\perp$ and, in the same range, the $C_p$ ratio for the $\beta=0.2$ case increases towards unity,
a {\em partial} contribution of KAW type of fluctuations - but not dominant, as highlighted by the $C_\|$ ratio, below - 
cannot be excluded in the low-$\beta$ regime.
Note that the decrease of $C_A$ at $k_\perp d_p\gg1$ is also consistent with a 
coupling with IB modes in both regimes~\citep{Groselj2017}.

Finally, let us consider the magnetic compressibility, $C_\|$ (Fig.~\ref{fig:SpectraRatio}, bottom panel): 
Alfv\'enic fluctuations would have small magnetic compressibility for $k_\perp\rho_p\ll1$
that increases as one goes to smaller and smaller scales and, in the KAW regime, eventually settles 
to a $\beta$-dependent value of $C_\|\simeq\beta_p/(1+2\beta_p)$ at $k_\perp\rho_p>1$ 
(represented in the bottom panel of Fig.~\ref{fig:SpectraRatio} by the green and blue horizontal dotted lines 
for $\beta_p=1$ and $\beta_p=0.2$, respectively).
Conversely, MS fluctuations have generally higher magnetic compressibility 
than the Alfv\'enic counterpart at $k_\perp\rho_p<1$ and, in the whistler regime, 
should settle to a $\beta$-independent value of  $C_\|=k_\perp/2k\lesssim1/2$ at $k_\perp~\rho_p~>~1$. 
From Fig.~\ref{fig:SpectraRatio} (bottom frame) we see that the magnetic compressibility
is consistent with Alfv\'enic/KAW-like fluctuations at $\beta_p=1$, i.e. it is small at $k_\perp\rho_p<1$
and then it increases to the nearly constant value of $C_\|\simeq\beta_i/(1+2\beta_i)=1/3$ 
expected for KAWs at $k_\perp\rho_p>1$.
The $\beta_i=0.2$ regime instead exhibits a magnetic compressibility
which is higher than that expected for Alfv\'enic/KAW fluctuations
throughout the whole $k_\perp$ range,
consistent with a mixture of MS, WWs and IB type of fluctuations~\citep{Groselj2017}.
Note that $k_\perp d_e$ effects can also enhance the compressibility of 
KAWs~\citep{ChenBoldyrev2017}, so, consistently with the previous ratios, 
there could be a non-negligible contribution of KAW-like fluctuations 
at $k_\perp\rho_p\gg1$ also in this low-$\beta$ regime.
All these results are qualitatively in agreement with previous
analysis performed in 2D fully-kinetic and hybrid-kinetic 
simulations~\citep{CerriAPJL2016,CerriJPP2017,Groselj2017}.

\section{Conclusions}

We presented the first high-resolution simulations of 3D3V
hybrid-kinetic turbulence including electron inertia effects
(with $m_p/m_e=100$), ranging from 
MHD scales to (perpendicular) scales well below the ion gyroradius. 
Two plasma beta parameters have been investigated:
an ``intermediate'' $\beta_p=1$ regime and a ``low'' $\beta_p=0.2$ case.

In both regimes, the spectral properties of the sub-proton turbulent cascade, 
such as its power laws and spectral anisotropy, and the intermittent behavior 
of the fluctuations are in good agreement with solar-wind observations and 
with the picture of turbulence mainly concentrated within 2D sheet-like structures 
presented in \citet{BoldyrevAPJL2012}.
In particular, all the turbulent fluctuations show a sub-proton-scale anisotropy pattern
of the type $k_\|\sim k_\perp^{2/3}$ and, correspondingly, the magnetic energy spectrum 
exhibits power-laws in perpendicular and parallel wave numbers that are
 $k_\perp^{-8/3}$ and $k_\|^{-7/2}$ at $\beta_p=1$
(being slightly steeper in $k_\perp$ and much more steeper in $k_\|$ 
for the low-$\beta$ case, roughly going as $k_\perp^{-3}$ and $k_\|^{-5}$).
This scenario has been supported also by intermittent analysis, which revealed 
deviations from monofractality and a strongly intermittent behavior at the kinetic scales
(the $\beta_p=0.2$ regime being slightly more intermittent than the intermediate-$\beta$ case).

Moreover, we find that the turbulent cascade is dominated by Alfv\'enic/KAW 
type of fluctuations at $\beta_p=1$, whereas the low-$\beta$ case presents a more
complex scenario suggesting the simultaneous presence of 
different types of fluctuations, including magnetosonic and whistler-like ones.
This picture seems to be supported also by the differences in the
parallel correlation length of the magnetic fluctuations between the two regimes, 
thus possibly indicating a different decorrelation mechanism along the mean field. 
Nevertheless, signatures that may be interpreted as 
ion Bernstein modes emerge in both regimes, 
although further focused investigations are needed in order to clarify this point.
The presence of IB fluctuations would indeed point to a link between kinetic turbulence, dissipation and 
reconnection~\citep{PodestaJGR2012,NaritaAnGeo2016}, as suggested also by 
the spectral properties~\citep{BoldyrevAPJL2012,LoureiroBoldyrev2017b,Mallet2017b}.

The results presented here are in qualitative agreement with previous
two-dimensional studies performed with fully-kinetic and hybrid-kinetic 
simulations~\citep{CerriAPJL2016,CerriJPP2017,Groselj2017}, 
although we stress that this scenario needs to include other important effects, 
such as the role of magnetic reconnection and the coupling with 
coherent structures~\citep{CerriCalifanoNJP2017,Franci2017}.
While the hybrid-kinetic model does not include all the electron kinetic physics
and larger resolutions would be needed to better separate the electrons and protons
kinetic scales, i.e., with a realistic mass ratio, the results presented here 
have a far-reaching implications in the context of solar-wind turbulence, 
from a possible dependence of the kinetic-scale cascade on the plasma $\beta$ parameter
to the understanding of the fundamental processes at play in collisionless kinetic plasma turbulence.

\acknowledgments

The authors acknowledge valuable discussions with 
F.~Pegoraro, L.~Franci, S.~Landi, E.~Papini, D.~Groselj and C.~H.~K.~Chen.
S.S.C. and F.C. thank C.~Cavazzoni (CINECA, Italy) 
for his essential contribution to the HVM code parallelization and performances.
The simulations were performed at CINECA (Italy)
under the ISCRA initiative (grant HP10BEANCY).

\bibliographystyle{apj}


\end{document}